\documentclass{article}

\newcommand{\nd}{\noindent}
\usepackage{graphics}
\usepackage[margin=2cm]{geometry}
\usepackage{graphicx}
\usepackage{amsmath}
\usepackage{amssymb}
\usepackage{amstext}
\usepackage{amscd}
\usepackage{amsfonts}
\usepackage{xcolor}
\title{Thermodynamically consistent entropic late-time cosmological acceleration}
\author{\small{D. J. Zamora$^{1,2}$\thanks{E-mail: javierzamora055@gmail.com}, C. Tsallis$^{1,3,4}$\thanks{E-mail: tsallis@cbpf.br}}, \\
\small{$^1$ Centro Brasileiro de Pesquisas Fisicas and}\\
\small{ National Institute of Science and Technology for Complex Systems,}\\
\small{ Rua Dr. Xavier Sigaud 150, Rio de Janeiro, 22290-180, Brazil.}\\
\small{$^2$ Instituto de Fisica del Noroeste Argentino,}\\
\small{ Av. Independencia 1800, Tucuman, CP 4000, Argentina}\\
\small{$^3$ Santa Fe Institute, 1399 Hyde Park Road, Santa Fe, 87501, NM, United States}\\
\small{$^4$  Complexity Science Hub Vienna, Josefstadter Strasse 39, Vienna, 1080, Austria}}

\date{\today}

\begin{document}

\maketitle
\begin{abstract}
Entropic-force cosmology provides, in contrast with dark energy descriptions, a concrete physical understanding of the accelerated expansion of the universe. The acceleration appears to be a consequence of the entropy associated with the information storage in the universe. Since these cosmological models are unable of explaining the different periods of acceleration and deceleration unless a correction term is considered, we study the effects of including a subdominant power-law term within a thermodynamically admissible entropic-force model. The temperature of the universe horizon is obtained by a clear physical principle, i.e., requiring that the Legendre structure of thermodynamics is preserved. We analyze the various types of behaviors, and we compare the performance of thermodynamically consistent entropic-force models with regard to available supernovae data by providing appropriate constraints for optimizing alternative entropies and temperatures of the Hubble screen. The novelty of our work is that the analysis is based on a entropy scaling with an arbitrary power of the Hubble radius, instead of a specific entropy. This allows us to conclude on various models at once, compare them, and conserve the scaling exponent as a parameter to be fitted with observational data, thus providing information about the form of the actual cosmological entropy and temperature. We show that the introduced correction term is capable of explaining different periods of acceleration and deceleration in the late-time universe.
\end{abstract}

\section{Introduction}
\label{sec:1}

The Lambda Cold Dark Matter ($\Lambda$CDM) model assumes a cosmological constant $\Lambda$ and the existence of dark energy. This model is the simplest one that can explain an accelerated expansion of the late universe. However, it implies several theoretical peculiarities, such as the cosmic coincidence and the cosmological constant problem \cite{Weinberg1989,Carroll2001}.
In mainstream cosmology, matter and space-time emerged from a singularity and evolved through four distinct periods, namely, early inflation, radiation, dark matter, and late-time expansion (driven by dark energy according to the $\Lambda$CDM model). During the radiation- and dark-matter-dominated stages, the universe is decelerating while the early and late-time expansion are accelerating stages. A possible connection between the accelerating periods remains unknown, and, intriguingly enough, the most popular dark energy candidate powering the present accelerating stage ($\Lambda$-vacuum) relies on the cosmological constant and coincidence puzzles. In order to handle these difficulties, several alternative models have been proposed, see for instance \cite{Weinberg2008,Ellis2012,Komatsu2014,Sola2013}. 

An interesting alternative model based on the concept of entropic-force is able to explain the late-time accelerated expansion of the universe \cite{Easson2011,Easson2012}. From this standpoint, the controversial dark-energy component is not necessary. Here, the late-time accelerated expansion is based on the entropic-force concept. Instead of the dark energy, we have the holographic principle and entropy as the source of the late accelerating phase of the universe. An entropic-force is an emergent phenomenon resulting from the natural tendency of a thermodynamical system to extremize its entropy, rather than from a particular underlying fundamental force. There is no field associated with an entropic-force. The force equation is expressed in terms of a spatial dependence of the entropy $S$. The cosmological entropic-force $F$, is then given by

\begin{equation}
F=-T\frac{dS}{dr_H},
\label{F}
\end{equation}

\nd where $r_H$ is the Hubble radius.

At this point, let us make an important clarification. The present entropic-force cosmological model is definitively different from the idea that gravity itself is an entropic-force, as suggested in \cite{Verlinde2011}. The entropic-force term has the potential of explaining the accelerated expansion without introducing new fields nor dark energy.

Thermodynamical properties of the universe have always attracted attention \cite{Davies1987,Prigogine1988,Prigogine1989} and, in more recent years, entropic cosmology in particular \cite{Cai2010,Karami2011,Mitra2015,Viaggiu2015}.
The first entropic-force model proposed by Easson, Frampton, and Smoot (EFS) \cite{Easson2011} assumes that the entropy and temperature associated to the horizon of the universe are the Bekenstein-Hawking entropy \cite{Bekenstein1973} and the Hawking temperature \cite{Hawking1974}, respectively. After that, other entropies were considered, such as the nonadditive $S_{\delta=3/2}$-entropy \cite{Komatsu2013}. This entropy was proposed in \cite{Tsallis2013} in the context of black-holes. Let us remind the reader that, for such systems, the additive Bekenstein-Hawking entropy is proportional to the area of the horizon, i.e. it is subextensive, whereas the nonadditive $S_{\delta=3/2}$-entropy is proportional to the volume (at least in the case of equal probabilities), i.e. it is extensive as required by thermodynamics. In the models \cite{Easson2011,Easson2012,Komatsu2013}, the expression of the temperature of the Hubble horizon is not reobtained from a neat physical principle. It is simply assumed to be the Hawking temperature expressed in terms of the universe parameters, namely

\begin{equation}
T_{BH}(t)=\frac{\hbar c}{2\pi k_Br_H(t)}=\frac{\hbar H(t)}{2\pi k_B},
\label{Thawking}
\end{equation}

\nd where $c$ is the speed of light, $\hbar$ is the reduced Planck constant, $k_B$ the Boltzmann constant, and $H(t)$ is the Hubble parameter. $H$ is defined as 

\begin{equation}
H\equiv\frac{c}{r_H}=\frac{\dot{a}}{a},
\end{equation}

\nd $a=a(t)$ being the scale factor.

Arbitrary combinations of entropy and temperature might violate the Legendre structure of thermodynamics. This is not the case of the Bekenstein-Hawking entropy and the Hawking temperature, as proposed by EFS \cite{Easson2011}. This issue was recently discussed in \cite{Zamora2021}, where we proposed a physical principle for deducing the thermodynamically consistent temperature associated to each class of entropy. More precisely, we deduce the temperature from the Legendre structure of thermodynamics:

\begin{equation}
\begin{split}
G(V,T, p,\mu,...)&=U(V,T,p,\mu,...)-TS(V,T,p,\mu,...)\\
&+pV-\mu N(V,T,p,\mu,...)-...,
\end{split}
\label{Legendre}
\end{equation}

\nd where $T,p,\mu$ are the temperature, pressure, and chemical potential, and $U,S,V,N$ are the internal energy, entropy, volume, and the number of particles of the system, respectively. This implies that, as detailed in \cite{Tsallis2013}, in a Schwarzschild (3+1)-dimensional black hole, the relation

\begin{equation}
\theta=1-d
\label{key2}
\end{equation}

\nd must hold, where $d$ is the dimension ($S\propto L^d$), and $\theta$ is the corresponding exponent for the scaling of the temperature ($T\propto L^\theta$), $L$ being a characteristic linear dimension of the d-dimensional system. For the Bekenstein-Hawking entropy ($d=2$), the temperature depends on $L^{-1}$ as the Hawking temperature.

The entropic-force term, Eq. (\ref{F}), affects the background evolution of the late universe; in the present paper we do not focus on the inflation of the early universe. It has been shown that entropic-force models which include only $H^2$ terms are not able to describe on a single footing both decelerating and accelerating stages \cite{Perico2013,Komatsu2013b}. Indeed, Basilakos et al. \cite{Basilakos2012} have shown that the first Easson-Frampton-Smoot (EFS) entropic-force model (which only includes a $H^2$ term) does not describe properly both acceleration and deceleration cosmological regimes unless a $\dot H$ term is included as well.

In our previous work \cite{Zamora2021}, we showed that the entropic-force term in the acceleration equation of all thermodynamically consistent models are of the $H^2$-type, similarly to the entropic-force term in the EFS model \cite{Easson2011}. As a consequence, the deceleration parameter, currently noted $q$, is a constant and thermodynamically consistent models are unable of predicting different stages of acceleration and deceleration. Nevertheless, changes in the deceleration parameter can be smoothly introduced by including subdominant terms in the entropy of the horizon. For example, EFS considered a logarithmic correction term in Bekenstein entropy \cite{Easson2012}. With this additional term, the model predicts different periods of acceleration and deceleration. However, they do not compare this prediction with the available data, neither sufficiently analyze the consequences of this specific addition. Following along this line but on more general grounds, we explore here the effects of adding a power-law subdominant term in the entropy and, at the same time, we consider an arbitrary entropy scaling with the power $d$ of the length. From this generalized approach, we study how several entropic-force models accommodate a viable cosmology for late-times without the consideration of dark energy. To develop some physical intuition concerning subdominant entropic terms we may think, as an illustration on a classical fluid, in a certain amount of water flowing, until final arrest, on a horizontal planar glass. We may observe that, just before the final equilibrium, the liquid surface returns slightly back (thus slightly decreasing the entire perimeter) due to surface tension effects. By this, what we mean is that subdominant entropic terms are not rare in nature, though they are neglected in first approximations. Even further, they have physical meaning and their inclusion are physically motivated.

\section{Subdominant term}
\label{sec:2}

Let us consider an entropy that scales with some arbitrary positive power $d\in\mathbb{R^+}$ plus an additional term depending on a smaller power $0<\Delta<d$.

\begin{equation}
\frac{S}{k_B}=A\left(\frac{r_H}{L_P}\right)^d+E\frac{\left(r_H/L_P\right)^{\Delta}-1}{\Delta},
\label{Snu}
\end{equation}

\nd where $A$ and $E$ are dimensionless constants, and $L_P=\sqrt{\hbar G/c^3}$ is the Planck length. 
By expressing the subdominant term in this way, we recover the logarithmic correction presented in \cite{Easson2012} when we take $\Delta\rightarrow0$, since\\
$\lim_{\Delta\rightarrow0}\frac{x^\Delta-1}{\Delta}=\ln{x}$. Additionally, by taking $E=0$ we obtain the Bekenstein-Hawking entropy (d=2) and the $\delta=3/2$ entropy (d=3) as particular cases.

According to Eq. (\ref{key2}), the thermodynamically correct temperature must scale like $T\propto r_H^{1-d}$. Consequently, we use

\begin{equation}
T=\frac{T_P}{B}\left(\frac{r_H}{L_P}\right)^{1-d},
\end{equation}

\nd as the temperature of the Hubble horizon, where $B$ is a dimensionless factor, and $T_P=\sqrt{\hbar c^5/G k_B^2}$ is the Planck temperature. The entropic force is then given by

\begin{equation}
\begin{split}
F&\equiv-T\frac{dS}{dr_H}=-k_B\frac{d\, A}{B}.\frac{T_P}{L_P}\left[1+\frac{E}{d\,A}\left(\frac{r_H}{L_P}\right)^{\Delta-d}\right]\\
&\equiv-C_d F_P(1+D_{d,\Delta}H^{d-\Delta}),
\end{split}
\end{equation}

\nd where $F_P\equiv k_BT_P/L_P=c^4/G$ is the Planck force, $C_d\equiv d\,A/B$, and $D_{d,\Delta}\equiv E(L_p/c)^{d-\Delta}/(d\,A)$. Therefore, the entropic pressure in the Hubble surface is

\begin{equation}
p_F\equiv\frac{F}{4\pi r_H^2}=-\frac{C_dc^2}{4\pi G}H^2(1+D_{d,\Delta}H^{d-\Delta}).
\end{equation}

Note that, when $d=2$ with $\Delta\rightarrow0$, we obtain the $H^4$ correction term in \cite{Easson2012}, and when $D_{d,\Delta}=0$ we obtain the $H^2$-type models \cite{Easson2011,Zamora2021}. To obtain the Friedmann equations modified by $p_F$, we replace the effective pressure $p'=p+p_F$ in the acceleration equation

\begin{equation}
\frac{\ddot{a}}{a}=-\frac{4\pi G}{3}\left(\rho+\frac{3p'}{c^2}\right),
\label{acc}
\end{equation}

\nd thus arriving to

\begin{equation}
\frac{\ddot{a}}{a}=-\frac{4\pi G}{3}\left(\rho+\frac{3p}{c^2}\right)+C_dH^2(1+D_{d,\Delta}H^{d-\Delta}).
\label{accelerationeq}
\end{equation}

In eqs. (\ref{acc}) and (\ref{accelerationeq}), $\rho$ is the total energy density of the universe. Replacing now $p'$ in the continuity equation 

\begin{equation}
\dot{\rho}+3\frac{\dot{a}}{a}\left(\rho+\frac{p'}{c^2}\right)=0,
\end{equation}

\nd we obtain

\begin{equation}
\dot{\rho}+3\frac{\dot{a}}{a}\left(\rho+\frac{p}{c^2}\right)=\frac{3C_d}{4\pi G}H^3(1+D_{d,\Delta}H^{d-\Delta}).
\label{continuityeq}
\end{equation}

Now, we follow the procedure of \cite{Komatsu2013} to derive a modified Friedmann equation from eqs. (\ref{accelerationeq}) and (\ref{continuityeq}). Considering the generalized Friedmann and acceleration equations,

\begin{equation}
\left(\frac{\dot{a}}{a}\right)^2=\frac{8\pi G\rho}{3}+f(t),
\end{equation}
\begin{equation}
\frac{\ddot{a}}{a}=-\frac{4\pi G}{3}\left(\rho+\frac{3p}{c^2}\right)+g(t),
\label{generalizedacceleration}
\end{equation}

\nd one deduces

\begin{equation}
\dot{\rho}+3\frac{\dot{a}}{a}\left(\rho+\frac{p}{c^2}\right)=\frac{3}{4\pi G}H\left(-f(t)-\frac{\dot{f}(t)}{2H}+g(t)\right).
\label{generalizedcontinuity}
\end{equation}

As examined in \cite{Komatsu2013b}, assuming a non-adiabatic-like expansion of the universe, we can simplify the model by considering a dependence of the form $f(t)=\alpha [H(t)]^2$. By comparing Eq. (\ref{continuityeq}) with (\ref{generalizedcontinuity}), and Eq. (\ref{accelerationeq}) with (\ref{generalizedacceleration}), we obtain $\alpha=0$. Consequently, the Friedmann equation is

\begin{equation}
\left(\frac{\dot{a}}{a}\right)^2=\frac{8\pi G\rho}{3}.
\label{friedmanneq}
\end{equation}

The three main equations are (\ref{accelerationeq}), (\ref{continuityeq}), and (\ref{friedmanneq}), but only two of them are independent.

We obtain the solution under the assumption of a homogeneous, isotropic, and spatially flat universe. From eqs. (\ref{accelerationeq}), (\ref{continuityeq}), and (\ref{friedmanneq}), we obtain

\begin{equation}
\begin{split}
&\frac{2C_dD_{d,\Delta}H_0^{d-\Delta}+[2C_d-3(1+\omega)]\left(\frac{H_0}{H}\right)^{d-\Delta}}{2C_dD_{d,\Delta}H_0^{d-\Delta}+[2C_d-3(1+\omega)]}=\\
&\left(\frac{a}{a_0}\right)^{-\frac{(d-\Delta)}{2}[2C_d-3(1+\omega)]},
\label{solutionnu}
\end{split}
\end{equation}

\nd where $\omega\equiv\frac{p}{\rho\,c^2}$; $a_0$ and $H_0$ are the contemporary values of $a$ and $H$, respectively. 
Let us focus now on the simple case of non-relativistic matter-dominated universe, i.e. $\omega=0$ \cite{Weinberg2008}. A straightforward calculation yields the following explicit time-dependent solution:

\begin{equation}
\begin{split}
&(1+\Delta-d)C_dD_{d,\Delta}H_0^{d-\Delta}H_0(t-t_0)=\left[(2C_dD_{d,\Delta}H_0^{d-\Delta}+\right.\\
&\left.2C_d-3)\left(\frac{a}{a_0}\right)^{-\frac{(d-\Delta)}{2}(2C_d-3)}-2C_dD_{d,\Delta}H_0^{d-\Delta}\right]^{1-\frac{1}{d-\Delta}}\times\\
&{}_2F_1\left(1,1-\frac{1}{d-\Delta};2-\frac{1}{d-\Delta};\right.\\
&\left.1-\frac{2C_dD_{d,\Delta}H_0^{d-\Delta}+2C_d-3}{2C_dD_{d,\Delta}H_0^{d-\Delta}}\left(\frac{a}{a_0}\right)^{-\frac{(d-\Delta)}{2}(2C_d-3)}\right)-\\
&(2C_d-3)^{1-\frac{1}{d-\Delta}}\times\\
&{}_2F_1\left(1,1-\frac{1}{d-\Delta};2-\frac{1}{d-\Delta};-\frac{2C_d-3}{2C_dD_{d,\Delta}H_0^{d-\Delta}}\right).
\label{avsteq}
\end{split}
\end{equation}

\section{Results}
\label{sec:3}

\begin{figure}
\centering
\resizebox{0.45\textwidth}{!}{\includegraphics{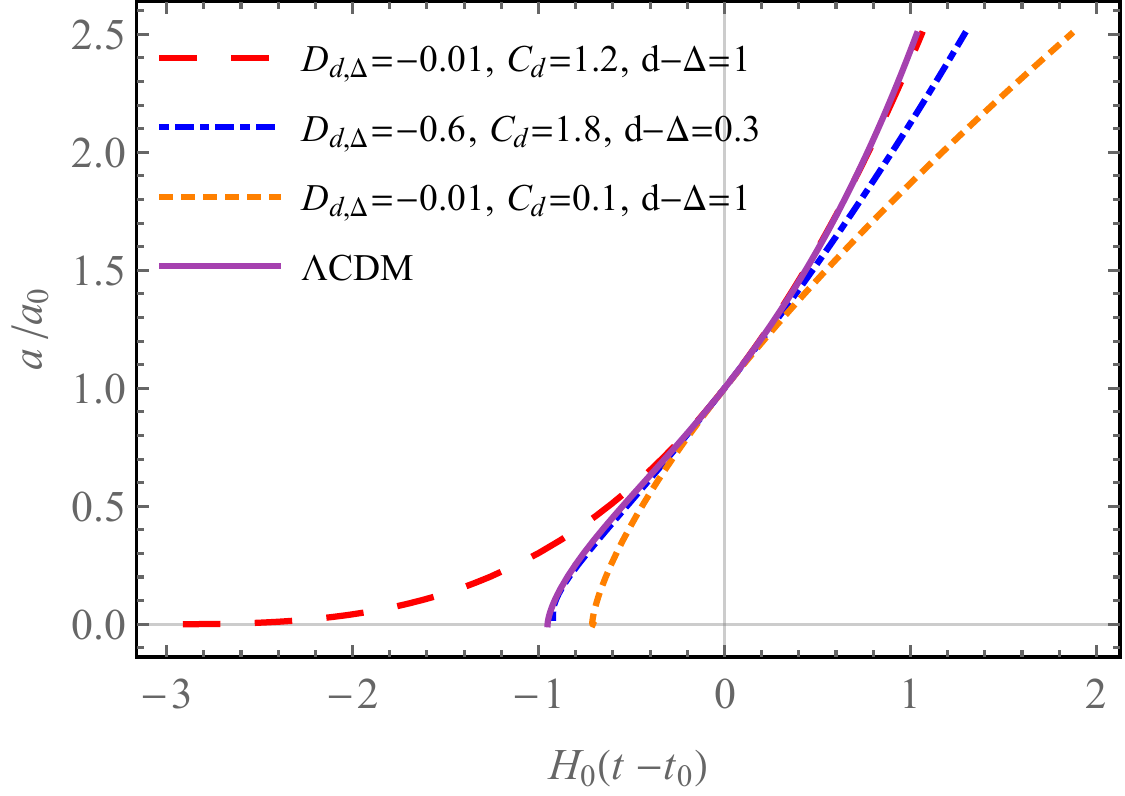}}
\caption{\label{avst} Time evolution of normalized scale factor $a/a_0$ for several combinations of the parameters. The horizontal axis is normalized as $H_0(t-t_0)$. $t_0$ and $a_0$ are the current values of time and scale factor, respectively. The possible behaviors of the solution are: i-convex curve (red dashed), ii-concave curve (orange dotted), and iii-curve with change of concavity (blue dot-dashed). The fine-tuned standard $\Lambda$CDM model (purple solid curve) is showed for comparison purposes.}
\end{figure}

In this section, we analyze the behavior of entropic-force models for late-time cosmology. To do so, in Fig. \ref{avst} we have plotted Eq. (\ref{avsteq}) for several combinations of the parameters, showing the different types of behaviors. We have plotted the solution for $C_d=1.2$, $D_{d,\Delta}=-0.01$, and $d-\Delta=1$ (red dashed curve) in which the curvature is always convex. The solution with parameters $C_d=0.1$, $D_{d,\Delta}=-0.01$, and $d-\Delta=1$ (orange dotted curve) is always concave. The blue dot-dashed curve has parameters $C_d=1.8$, $D_{d,\Delta}=-0.6$, and $d-\Delta=0.3$ and it presents a change of concavity, in a similar way to the fine-tuned standard $\Lambda$CDM model ($\Omega_m=0.315, \Omega_\Lambda=0.685$, purple solid curve). In other words, the entropic-force model with subdominant term is able to predict stages of decelerated and accelerated expansion of the universe, which is similar to the fine-tuned standard $\Lambda$CDM model. The blue curve is indistinguishable from the $\Lambda$CDM model for $t<t_0$, and the red curve is indistinguishable from the $\Lambda$CDM model for $t>t_0$.

Let us study how the subdominant term influences the behavior of the deceleration parameter, $q\equiv-\ddot a/(aH^2)$. We remind the reader that without a subdominant term, entropic-force models are not capable of accommodating a viable cosmology, since the deceleration parameter is always negative and, therefore, they predict a universe in perpetual accelerated expansion \cite{Zamora2021,Basilakos2012,Easson2011}. The deceleration parameter straightforwardly follows from Eq. (\ref{avsteq}), and is given by

\begin{equation}
\begin{split}
q&=-\frac{1}{2}(2C_d-3)-1-C_dD_{d,\Delta}H^{d-\Delta}\\
&=-\frac{1}{2}(2C_d-3)-1-\\
&\frac{C_dD_{d,\Delta}}{(z+1)^{\frac{1}{2}(d-\Delta)(2C_d-3)} \left(H_0^{-(d-\Delta)}+\frac{2C_dD_{d,\Delta}}{2C_d-3}\right)-\frac{2C_dD_{d,\Delta}}{2C_d-3}}.
\label{deceleration}
\end{split}
\end{equation}

\begin{figure*} 
\centering
  \begin{minipage}{0.49\linewidth}
  \centering
   \resizebox{0.95\textwidth}{!}{\includegraphics{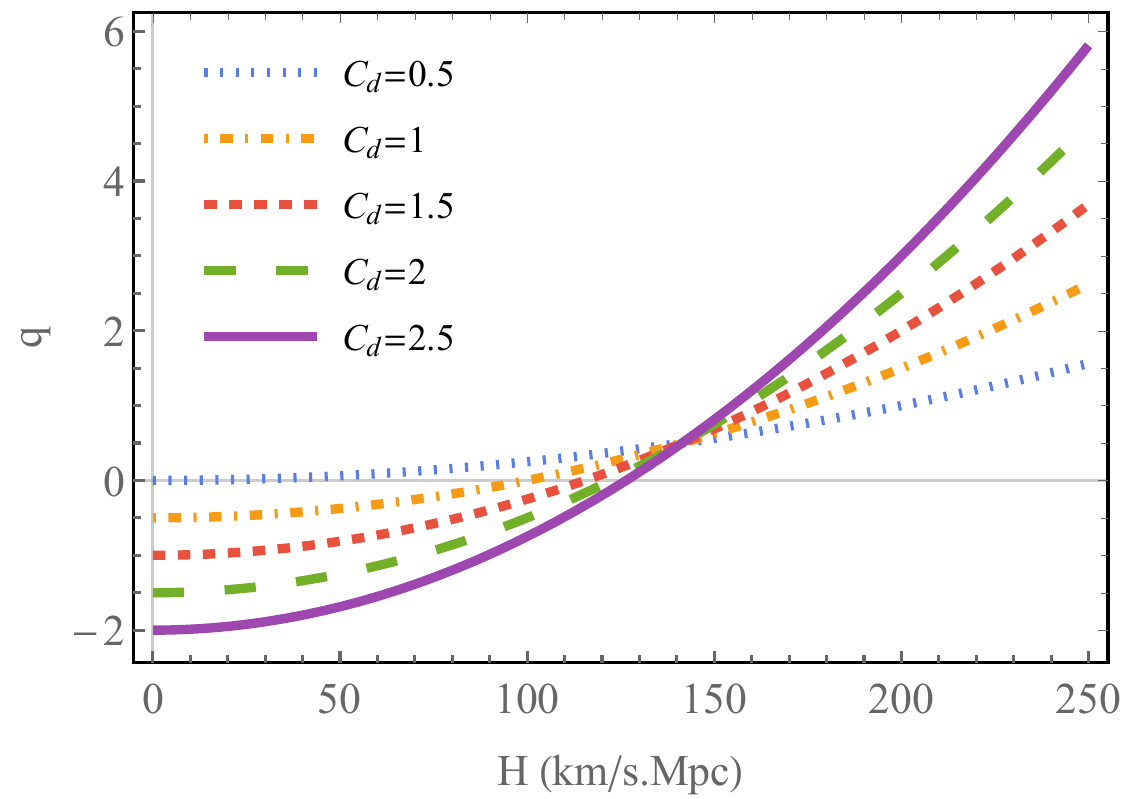}}
    \caption{\label{qc} Deceleration parameter $q$ versus Hubble parameter $H$ for five different values of $C_d$, $D_{d,\Delta}=0.00005$, and $d-\Delta=2$. Notice the change of sign.} 
  \end{minipage} 
  \hfill
  \begin{minipage}{0.49\linewidth}
  \centering
   \resizebox{0.95\textwidth}{!}{\includegraphics{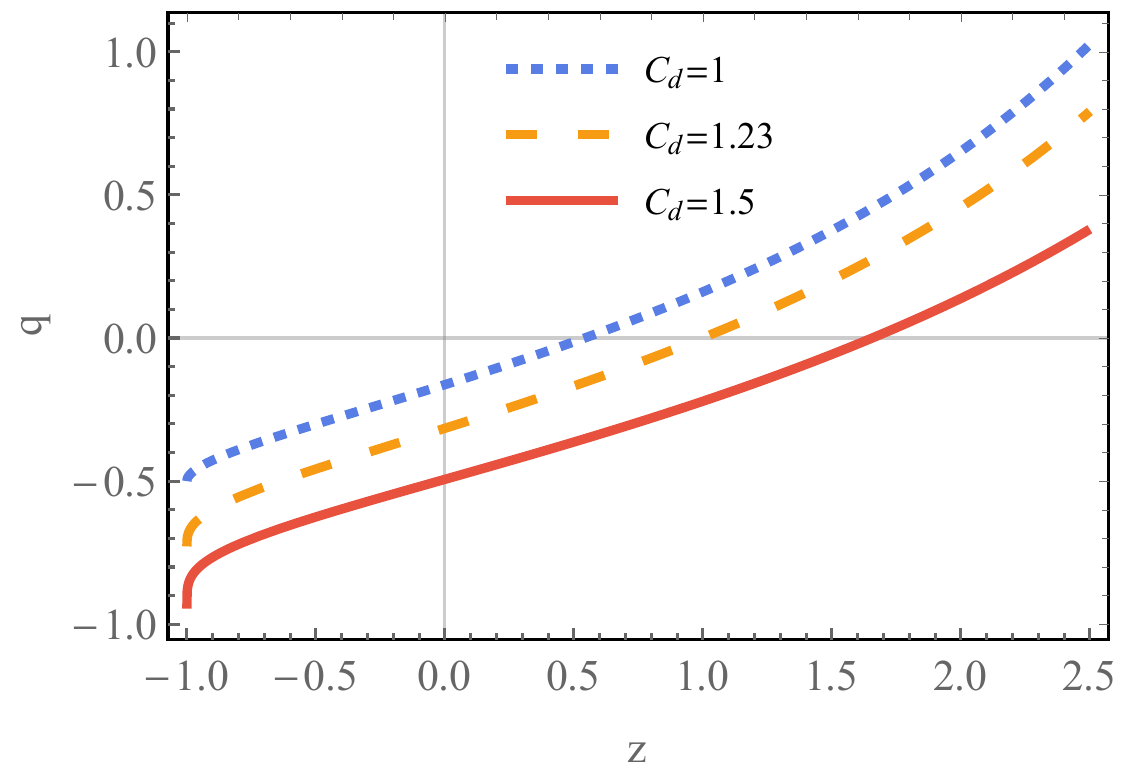}}
       \caption{\label{qvszc} Deceleration parameter $q$ versus redshift $z$ for three different values of $C_d$, $D_{d,\Delta}=-0.005$, and $d-\Delta=1$. $C_d$ changes the value of $q$ in $z=-1$.} 
  \end{minipage} 
  \\
   \begin{minipage}{0.49\linewidth}
    \centering
   \resizebox{0.95\textwidth}{!}{\includegraphics{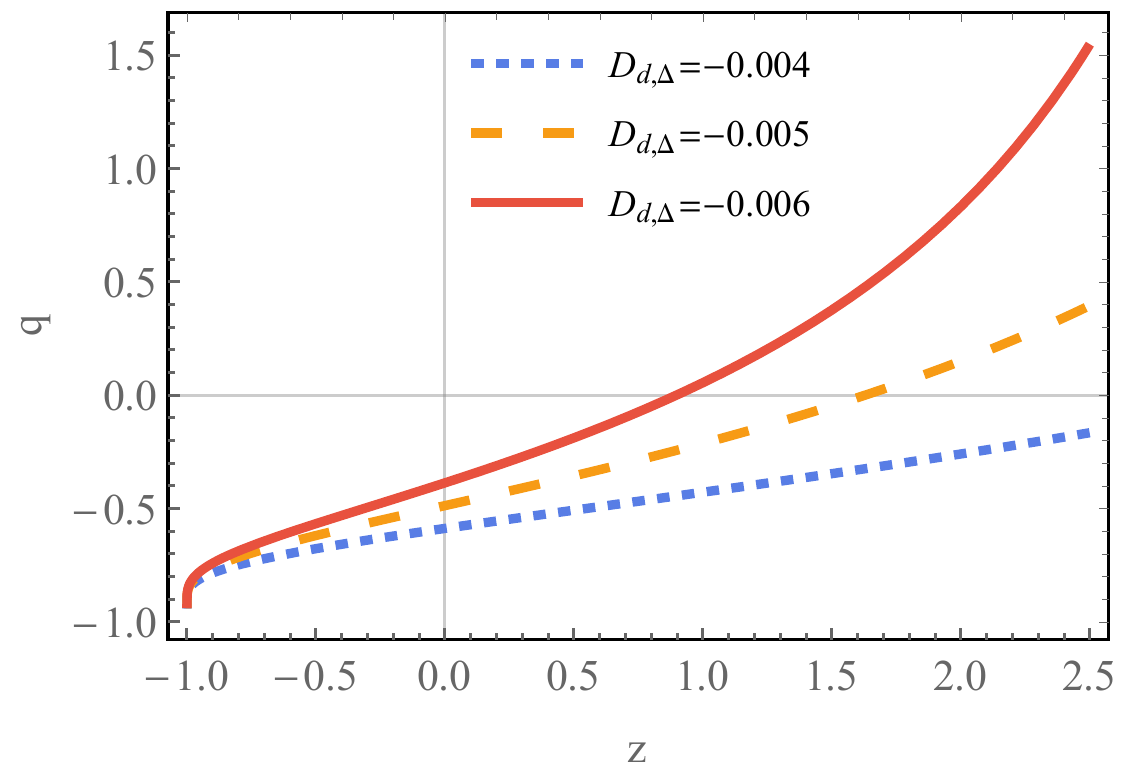}}
    \caption{\label{qvszd} Deceleration parameter $q$ versus redshift $z$ for three different values of $D_{d,\Delta}$, $C_d=1.5$, and $d-\Delta=1$.}
  \end{minipage} 
  \hfill
   \begin{minipage}{0.49\linewidth}
    \centering
   \resizebox{0.95\textwidth}{!}{\includegraphics{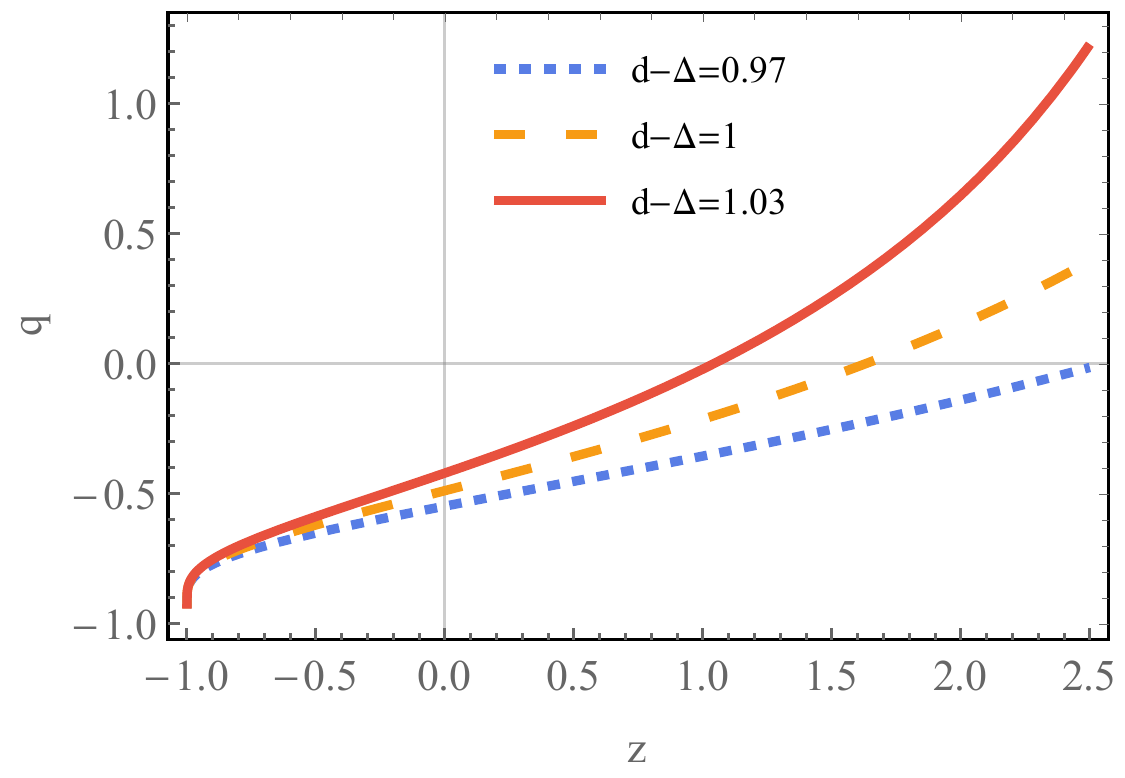}}
    \caption{\label{qvszDelta} Deceleration parameter $q$ versus redshift $z$ for three different values of $d-\Delta$, $C_d=1.5$, and $D_{d,\Delta}=-0.005$. Notice the big changes in $q$ for small variations in $d-\Delta$.}
  \end{minipage} 
\end{figure*}

The deceleration parameter does depend on $H$, and therefore, on time. The inclusion of a first-order correction to the horizon entropy provides a natural source of dependence of the deceleration parameter with the redshift $z$. The equation describing $H(z)$ is obtained by replacing the definition of the redshift, $1+z\equiv a_0/a$, in Eq. (\ref{solutionnu}). Values of $q<0$ correspond to an accelerating universe and $q>0$ to a decelerating one. Depending on the combination of parameters $C_d$, $D_{d,\Delta}$, $d$, and $\Delta$, the deceleration parameter is positive or negative, and it is able to explain periods of acceleration and deceleration, as shown in Figs. \ref{qc}-\ref{qvszDelta}. In Fig. \ref{qc} we plot the deceleration parameter for five different values of $C_d$. They intersect in the point $H=(-D_{d,\Delta})^{1/(d-\Delta)}$, and $q=1/2$. The physical interpretation of this interesting point remains elusive at the present stage. Let us also mention that $q=1/2$ is precisely the value of the deceleration parameter at $t=0$ in the $\Lambda$CDM model.

In Figs. \ref{qvszc}-\ref{qvszDelta} we have plotted the deceleration parameter as a function of redshift $z$ for recent times (low negative and positive values of $z$) for various values of ($C_d$, $D_{d,\Delta}$, $d-\Delta$). $C_d$ has the effect of changing the value of $q$ in the point $z=-1$. When $C_d=1.5$, $q=-1$, as in the $\Lambda$CDM model and others \cite{Sasidharan2015,Moradpour2020}. However, the shape of the curve is different from the $\Lambda$CDM one. Instead, it is similar to that reported in \cite{Tiwari2021}. In all cases, when $z=0$ (current time) $q<0$, thus recovering the knowledge that the universe is currently accelerating. Finally, notice the change of sign, meaning that the subdominant term in the horizon entropy is capable of explaining both decelerated and accelerated expansions. Unlike $\Lambda$CDM in which $q$ approaches $0.5$ when $z$ goes to infinity, $q$ grows without restriction for the early universe in this model. Then, one should restrict the present analysis to the late-time and close-future universe.

\begin{figure}
\centering
\resizebox{0.45\textwidth}{!}{\includegraphics{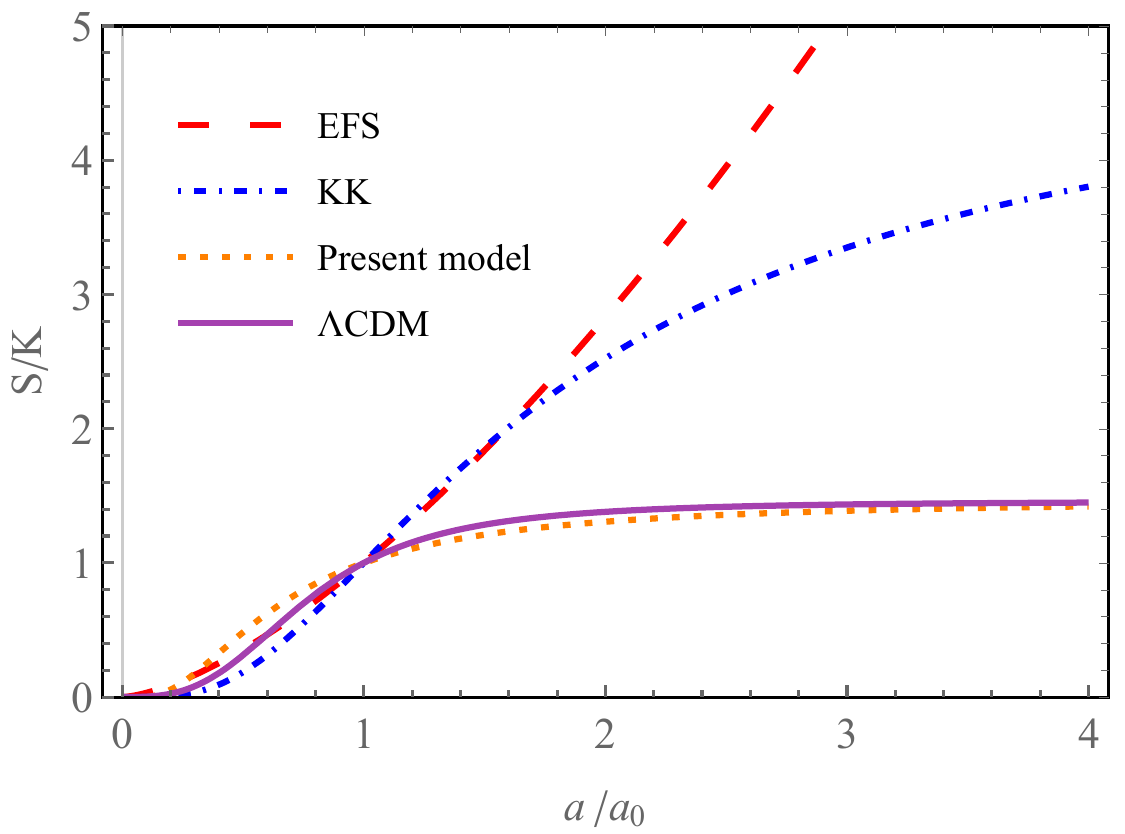}}
\caption{\label{entropy} Evolution of the Bekenstein-Hawking and generalized entropies. The vertical axis represents dimensionless entropy, where the parameter K is $K_{BH}=\pi k_B c^5/\hbar GH_0^2$ and $K_{KK}=\pi k_B c^6/\hbar GH_0^3$, for the Bekenstein-Hawking entropy used in EFS entropic model (EFS, red dashed curve) and generalized Komatsu-Kimura entropic-force model (KK, dot-dashed blue curve), respectively. The purple solid line represents $S_{BH}/K_{BH}$ for the fine-tuned standard $\Lambda$CDM model and it is analytically calculated from $S_{BH}/K_{BH}=H^{-2}=(a/\dot{a})^2$. $K=k_B A$ for our generalized model with power correction term (orange dotted curve). Depending on the parameters, our model presents diverse behaviors, ranging from curves similar to EFS model ($C_d=3,D_{d,\Delta}=-0.25,d=2,\Delta=1,E=0$), passing through Komatsu-Kimura model ($C_d=3,D_{d,\Delta}=-0.29465,d=3,\Delta=2,E=0$), until a curve similar to the $\Lambda$CDM model ($C_d=3.3,D_{d,\Delta}=-0.45,d=2,\Delta=1,E=0.01$), which is the plotted orange dotted curve.}
\end{figure}

Let us focus now on the entropy evolution. In Fig. \ref{entropy}, we show the behavior of different dimensionless entropies with the scale factor. The entropies are normalized in different ways (see caption). We analytically calculated the Bekenstein-Hawking entropy of the fine-tuned standard $\Lambda$CDM model as $S_{BH}/K_{BH}=H^{-2}=(a/\dot{a})^2$, ($K_{BH}\equiv\pi k_B c^5/\hbar GH_0^2$), in order to compare with the entropic-force models. For $a/a_0<1$, the entropy for the standard $\Lambda$CDM model increases rapidly, whereas, for $a/a_0>1$, the increment in the entropy tends to become gradually slower. Similar results have been reported in \cite{Easther1999,Barrow2003,Davies2003,Wang2006,Setare2006,Cline2008,Egan2010}. We now examine the entropic-force models. Let us emphasize at this point that, for $a/a_0<1$, the entropy for all entropic-force models is consistent with the standard $\Lambda$CDM model. However, for $a/a_0>1$, the entropy for the EFS entropic-force model increases uniformly, whereas the increment in the entropy for the generalized Komatsu and Kimura (KK) entropic-force model \cite{Komatsu2013} tends to become gradually slower. On the other hand, for $a/a_0>1$, the entropy for this generalized KK entropic-force model increases more rapidly than for the $\Lambda$CDM model. The evolution of the entropy for the present generalized entropic-force model with power-law subdominant term exhibits diverse behaviors depending on the parameters values, ranging from curves similar to the EFS model ($C_d=3,D_{d,\Delta}=-0.25,d=2,\Delta=1,E=0$), passing through the Komatsu-Kimura model ($C_d=3,D_{d,\Delta}=-0.29465,d=3,\Delta=2,E=0$), until eventually attaining curves similar to the $\Lambda$CDM model for all $a/a_0$. For instance, $C_d=3.3,D_{d,\Delta}=-0.45,d=2,\Delta=1,E=0.01$ correspond to the plotted orange dotted curve in Fig. \ref{entropy}.

\section{Comparison with supernova data}
\label{sec:4}

\begin{figure}
\centering
\resizebox{0.45\textwidth}{!}{\includegraphics{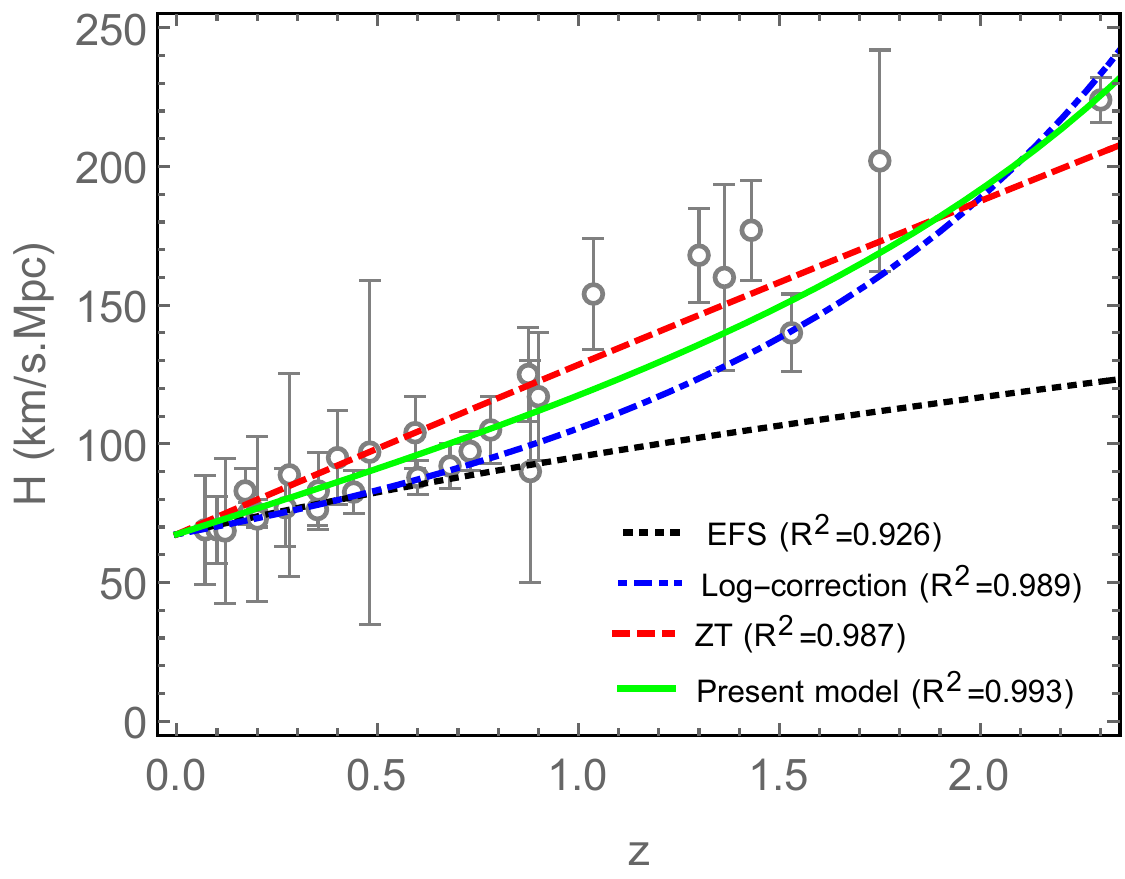}}
\caption{\label{Hvsz} Hubble parameter $H$ versus redshift $z$. The open circle with bars are data points taken from Table 1 in \cite{Pradhan2021}. In all cases, the value of $H_0$ is set to be $67.4 km/s/Mpc$ based on the Planck 2018 results \cite{PlanckCollaboration2018}. The black dotted curve is the first EFS entropic-force model \cite{Easson2011} ($d=2$, $\Delta\rightarrow0$, $D_{d,\Delta}=0$, $C_d$=1). The dot-dashed blue curve is the second EFS model \cite{Easson2012} with logarithmic correction term ($d=2$, $\Delta\rightarrow0$, $C_d=1$). The fitting value is $D_{d,\Delta}=(0.1322\pm0.0004)\times10^{-3}$. The dashed red curve is our previous thermodynamically consistent entropic-force model \cite{Zamora2021}
($D_{d,\Delta}=0$). The fitting value is $C_d=0.57\pm0.03$. The solid green curve is our present model with power correction. The best fitting values are $C_d=1.23\pm0.03$, $D_{d,\Delta}=(-0.005\pm0.001)$ and $d-\Delta=1.0\pm0.1$.}
\end{figure}

Supernova data are the main source of available measurements in order to compare cosmological models. They constitute nowadays one of the best observational tools for comparing several entropic-force models. We present here an analysis of the available updated data. In Fig. \ref{Hvsz}, we have plotted the Hubble parameter $H$ as a function of the redshift $z$ using the data points taken from table 1 in \cite{Pradhan2021}. We have plotted the best fittings for four different entropic-force models. In all cases, the value of $H_0$ is set to be $67.4\,(km/s)/Mpc$ based on the Planck 2018 results \cite{PlanckCollaboration2018}. The first EFS entropic-force model \cite{Easson2011} (black dotted curve) uses the Bekenstein-Hawking entropy and the Hawking temperature. This is a particular case of our present model ($d=2$, $D_{d,\Delta}=0$, $C_d$=1). Log-correction corresponds to the second EFS model \cite{Easson2012} (dot-dashed blue curve), which includes a logarithmic subdominant term. This particular case can be obtained from our model by taking $d=2$, $\Delta\rightarrow0$, $C_d=1$. We fitted $D_{d,\Delta}=(0.1322\pm0.0004)\times10^{-3}$. ZT stands for our previous thermodynamically consistent entropic-force model (dashed red curve) \cite{Zamora2021}.
In this case the entropy scales with an arbitrary power $d$, and can be obtained by taking $D_{d,\Delta}=0$. The fitting value is $C_d=0.57\pm0.03$. For our current entropic-force model with power correction (solid green curve), $C_d=1.23\pm0.03$, $D_{d,\Delta}=-0.005\pm0.001$ and $d-\Delta=1.0\pm0.1$ are determined by optimally fitting the data points. Notice that $d-\Delta$ is the difference of two dimensions ($d>\Delta$), therefore a positive integer number is welcome.

In addition to the above, we obtain a satisfactory agreement for the luminosity distance data points $d_L$ for all models, using the above fitting parameters. The luminosity distance is defined (see \cite{Komatsu2013,Easson2011} for instance) by

\begin{equation}
d_L(z)\equiv\frac{c(1+z)}{H_0}\int_1^{1+z}\frac{dy}{F(y)},
\end{equation}

\nd where $y\equiv a_0/a$, and $F(y)\equiv H(y)/H_0$. We remind that we are assuming $\omega=0$. From Eq. (\ref{solutionnu}), we obtain

\begin{equation}
\begin{split}
&\frac{H_0}{c}d_L=\frac{(1+z)(2C_d-3)}{2C_dD_{d,\Delta}H_0^{d-\Delta}}\times\left\{{}_2F_1\left[1,1+\frac{2C_d-1}{(2C_d-3)(d-\Delta)},\right.\right.\\
&\left.\left.1+\frac{2}{(2C_d-3)(d-\Delta)},1+\frac{2C_d-3}{2C_dD_{d,\Delta}H_0^{d-\Delta}}\right]-(1+z)\times\right.\\
&\left.\left[\frac{(2C_d-3+2C_dD_{d,\Delta}H_0^{d-\Delta})(1+z)^{\frac{d-\Delta}{2}(2C_d-3)}}{2C_d-3}\right.\right.\\
&\left.\left.-\frac{2C_dD_{d,\Delta}H_0^{d-\Delta}}{2C_d-3}\right]^{1+\frac{1}{d-\Delta}}\times\right.\\
&\left.{}_2F_1\left[1,1+\frac{2C_d-1}{(2C_d-3)(d-\Delta)},1+\frac{2}{(2C_d-3)(d-\Delta)},\right.\right.\\&\left.\left.\frac{(2C_d-3+2C_dD_{d,\Delta}H_0^{d-\Delta})(1+z)^{\frac{d-\Delta}{2}(2C_d-3)}}{2C_dD_{d,\Delta}H_0^{d-\Delta}}\right]\right\}.
\label{dLnu}
\end{split}
\end{equation}

\begin{figure}
\centering
\resizebox{0.45\textwidth}{!}{\includegraphics{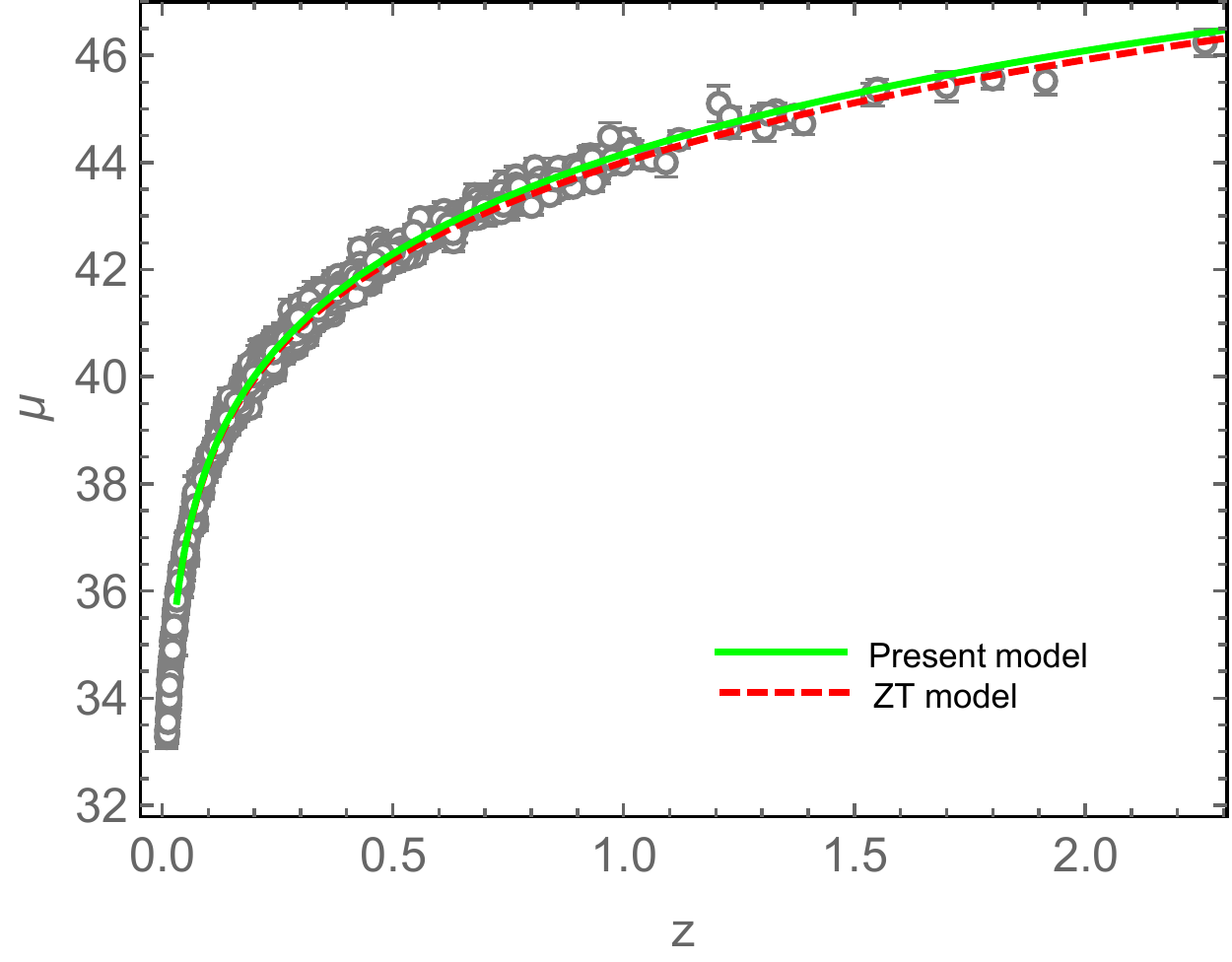}}
\caption{\label{dL2} Dependence of the distance modulus $\mu$ with redshift $z$. The open circles with error bars are supernova data points taken from \cite{Scolnic2018}.  SN1a absolute magnitude $M_0=-19.36$. The dashed red curve is our previous thermodynamically consistent model \cite{Zamora2021}. The solid green curve is the present model with power correction term. All entropic force models have a good agreement, with $R^2=0.999...$, we plotted two of the models for simplicity. In both cases, $H_0= 67.4\,(km/s)/Mpc$.}
\end{figure}

In Fig. \ref{dL2}, we have plotted the distance modulus $\mu$ versus redshift $z$ data taken from the so-called ”Pantheon Survey”, consisting of a total of 1048 Type Ia Supernovae \cite{Scolnic2018}, where

\begin{equation}
\mu=5\log_{10}d_L-5,
\end{equation}

\nd with $d_L$ in parsec. Figure \ref{dL2} displays the Pantheon Survey as the standard Hubble diagram of SN1a (absolute magnitude $M_0=-19.36$). We plotted our first model (ZT model, red dashed curve) as well as the current model with subdominant term (solid green curve). All entropic-force models fit similarly the modulus distance data points; we only plotted two of them as illustrations. In fact, there exist numerous and diverse cosmological models exhibiting a similar agreement with such data (see, for example, \cite{Komatsu2014,Tartaglia2009,Knop2003,Jana2014,Zhang2008}).

More complete data analyzes (for example using covariance matrix) are needed to further validate these models but this is out of the present scope. Here, we only used the data to compare the performance of different entropic-force models.

\begin{figure}
\centering
\resizebox{0.45\textwidth}{!}{\includegraphics{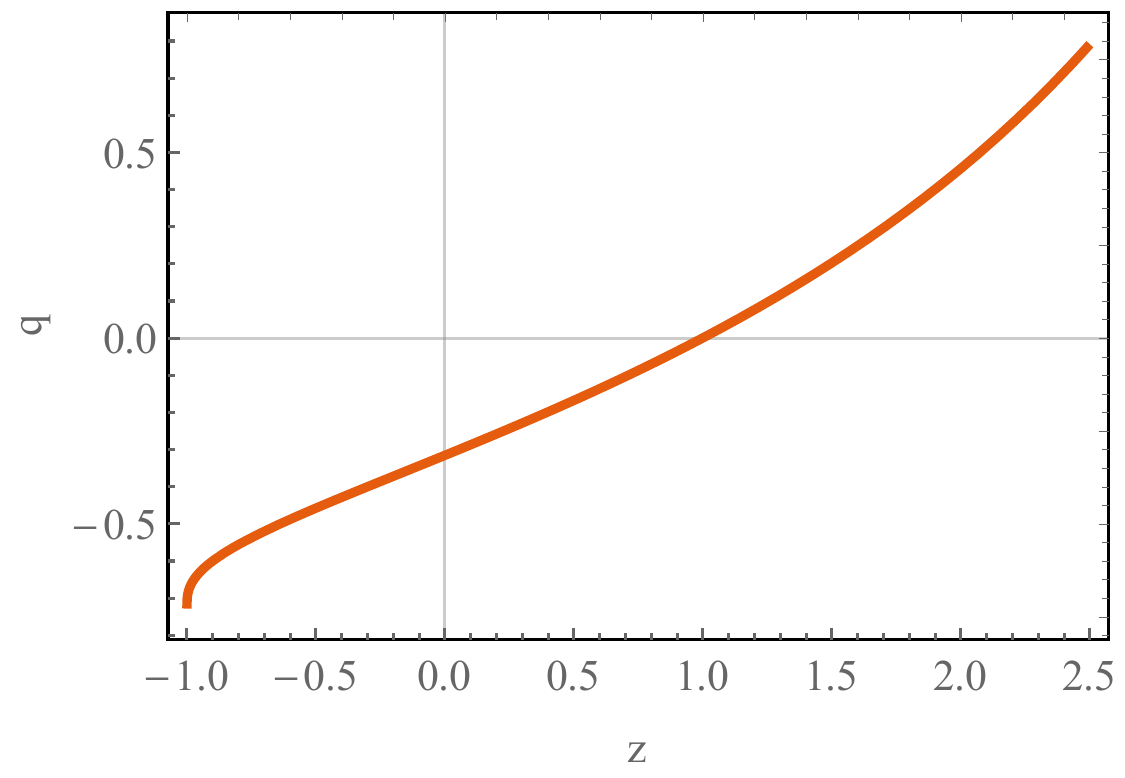}}
\caption{\label{qvsz} Dependence of the deceleration parameter $q$ with redshift $z$ for the fitted values $C_d=1.23$, $D_{d,\Delta}=-0.005$, and $d-\Delta=1$. $H_0$ was set to be $67.4\,(km/s)/Mpc$. See Eq. (\ref{deceleration}).}
\end{figure}

Finally, in Fig. \ref{qvsz} we have plotted the deceleration parameter $q$ as a function of the redshift $z$ with the fitting parameters for the present model with subdominant power-law term. Notice the change of sign. This means that the fitting parameters are coherent with stages of accelerated and decelerated expansion for the late-time expansion.

\section{Conclusions}
\label{sec:5}

Summarizing, in order to examine the entropic cosmology, we have assumed a generalized entropy with including a power-law subdominant term in the cosmological equations. This approach provides, in contrast with the dark energy description, a concrete physical understanding of the acceleration. The current accelerated expansion rate is the inevitable consequence of the entropy associated with the information storage in the universe. The different periods of acceleration and deceleration can be explained by such a correction term.

The choice of the entropy and temperature of the horizon of the universe is based on the Legendre structure of thermodynamics. It is on this basis that we have formulated the modified Friedmann, acceleration, and continuity equations. We showed that the Friedmann equation itself does not include the entropic-force term, in variance with the acceleration equation which presents the $H^2$ term as well as a $H^{d-\Delta+2}$ correction term, and the continuity equation which presents the $H^3$ term as well as a $H^{d-\Delta+3}$ correction term.

We have obtained a solution of the model, assuming a homogeneous, isotropic, and spatially flat universe. We have confirmed that this model describes a currently accelerating universe, without adding a cosmological constant or assuming the existence of dark energy. The power-law correction term in the entropy is capable of explaining the periods of acceleration and deceleration since the deceleration parameter can take both positive and negative values. In fact, the solution has three types of behaviors, namely, (i) an always accelerated expanding universe, (ii) an always decelerated expanding universe, and (iii) a decelerating and accelerating expanding universe similar to the $\Lambda$CDM model. However, the shape of the curve is different from the $\Lambda$CDM one. Instead, it is similar to that reported in \cite{Tiwari2021}. Unlike $\Lambda$CDM in which $q$ approaches $0.5$ when $z$ goes to infinity, $q$ grows without restriction for the early universe in this model. Then, one should restrict the present analysis to the late-time and close-future universe. In addition, depending on the values chosen for the parameters of the model, the behavior of the entropy is similar to that of the $\Lambda$CDM model. 

Finally, we compared the performance of diverse entropic force models with regard to the available supernova data. They show satisfactory agreement with the distance luminosity, and the Hubble parameter $H$ as a function of redshift $z$. 

To sum up, this kind of models is capable of predicting stages of deceleration and acceleration in the late cosmic expansion, a time-evolution of entropy similar to mainstream cosmology, being consistent with data. In addition, the fitting values of the parameters are consistent with a deceleration parameter changing signature around $z=1$.

As a serious alternative to mainstream cosmology, entropic force models need to satisfactory handle three important points: (i) validation through the full data analysis, including covariance matrices; (ii) correct explanation of the different periods of acceleration and deceleration; and (iii) a physical principle that mandates the entropy and temperature to be used for the Hubble horizon. In the present paper we have focused on the latter two points.

\section*{Acknowledgement}
This work has received financial support from CAPES, CNPq, and FAPERJ (Brazilian agencies).

\end{document}